\documentclass[12pt]{article}
\usepackage{fullpage}
\usepackage{amssymb,amsmath,amstext,amscd,mathrsfs,eucal,bm}

\usepackage[square,numbers]{natbib}
\usepackage{lineno}
\usepackage{amsfonts,amsbsy}
\usepackage{graphicx}
\usepackage{color}

\usepackage{url}            

\usepackage[percent]{overpic}
\usepackage{xcolor}
\usepackage{pict2e}

\usepackage{mathtools}
\usepackage{color}
\usepackage{tikz}
\usepackage{tkz-euclide} 

\newcounter{subeq}

\DeclareMathAlphabet\mathbfcal{OMS}{cmsy}{b}{n}

\newcommand{\bs}[1]{\boldsymbol{#1}}

\def\e{\mathrm{e}}

\def\d{\mathrm{d}}

\def\eps{\varepsilon}

\def\bn{\boldsymbol{n}}
\def\bx{\boldsymbol{x}}
\def\bX{\boldsymbol{X}}

\def\br{\boldsymbol{r}}

\def\br{\boldsymbol{r}}

\def\bxi{\boldsymbol{\xi}}

\def\bp{\boldsymbol{p}}

\def\bX{\boldsymbol{X}}

\def\radius{a}

\newcommand{\Abs}[1]{\ensuremath{\left| #1 \right|}}
\def\dd{{\rm d}} 

\def \p{{\bf p}}
\def \n{{\bf n}}
\def \x{{\bf x}}

\newcommand{\JV}[1]{\textcolor{black}{#1}}

\def\XXint#1#2#3{{\setbox0=\hbox{$#1{#2#3}{\int}$}
\vcenter{\hbox{$#2#3$}}\kern-.5\wd0}}

		\def\beq{\begin{equation}}
		\def\eeq{\end{equation}}
		\newcommand{\eref}[1]{(\ref{eqn:#1})}
		\newcommand{\elab}[1]{\label{eqn:#1}}
		
		\newcommand{\fref}[1]{\ref{fig:#1}}
		\newcommand{\flab}[1]{\label{fig:#1}}
		
		\newcommand{\sref}[1]{\ref{sec:#1}}
  		\newcommand{\slab}[1]{\label{sec:#1}}

\def\nw#1{\textcolor{black}{#1}}

\title{Diffusion in arrays of obstacles: beyond homogenisation}

\author{Yahya Farah$^1$, Daniel Loghin$^1$, Alexandra Tzella$^1$\thanks{Corresponding author: a.tzella@bham.ac.uk} and Jacques Vanneste$^2$\smallskip  \\
\normalsize{$^1$School of Mathematics, University of Birmingham, Birmingham, UK}, \\ \normalsize{$^2$ School of Mathematics and Maxwell Institute for Mathematical Sciences,} \\ \normalsize{University of Edinburgh, Edinburgh, UK}}

 \date{\today}

\begin{document}

\maketitle

\begin{abstract}
We revisit the classical problem of diffusion of a scalar (or heat) released in a two-dimensional medium with an embedded periodic array of impermeable obstacles such as perforations. Homogenisation theory provides a coarse-grained description of the scalar at large times and predicts that it diffuses with a certain effective diffusivity, so the concentration is approximately Gaussian. We improve on this by developing a large-deviation approximation which also captures the non-Gaussian tails of the concentration through a rate function obtained by solving a family of eigenvalue problems. We focus on cylindrical obstacles and on the dense limit,  when the obstacles occupy a large area fraction and non-Gaussianity is most marked.  We derive an asymptotic approximation for the rate function in this limit,  valid uniformly over a wide range of distances. We use finite-element implementations to solve the eigenvalue problems yielding the rate function for arbitrary obstacle area fractions and an elliptic boundary-value problem arising in the asymptotics calculation. Comparison between numerical results and asymptotic predictions confirm the validity of the latter.

\end{abstract}

\section{Introduction}

We consider the diffusion of a passive scalar inside a two-dimensional homogeneous medium interrupted by an infinite number of impermeable obstacles (e.g., perforations) arranged in a periodic lattice, as illustrated in Fig.\ \fref{problem_geometry} for the case of circular obstacles.
The scalar concentration $\theta(\bx,t)$ satisfies the dimensionless diffusion equation
\begin{subequations}\elab{diff}
\beq\elab{diff-eqn}
\frac{\partial \theta}{\partial t}=\nabla^2 \theta,
\eeq
with no-flux conditions on the boundaries $\mathcal{B}$ of the obstacles,
\beq\elab{noflux}
\bn\cdot\nabla \theta=0 \quad \textrm{on} \ \ \mathcal{B},
\eeq
\end{subequations}
where $\bn$ denotes the outward  normal to  $\mathcal{B}$.
The  concentration  is a function of the dimensionless position vector $\bx=(x,y)^\mathrm{T}$ scaled by a reference length scale $\ell$
related to the lattice period,  and time $t$ scaled by the diffusive timescale $\ell^2/\kappa$, where $\kappa$ is the molecular diffusivity.

We are interested in the initial-value problem corresponding to the instantaneous release of the scalar at some location  $\bx_0$ outside the obstacles.
 Our aim is to provide a coarse-grained description of $\theta(\bx,t)$, valid when the scalar has spread over many periods of the lattice. This problem and its steady-state counterpart have a long history, dating back to  Maxwell \cite{Maxwell1873} and Rayleigh \cite{Rayleigh1892}, driven by 
 \nw{the} relevance \nw{of these problems} to a broad range of applications that include constituent dispersion, heat conduction (with $\theta$ the temperature) and (with suitable re-interpretation)  electric conduction and electrostatics, in porous media and in composite materials (see e.g.\ Ch.\ 2 of \cite{Berlyand2012} for a survey). The central conclusion is that coarse-graining results in a diffusion equation,
\beq \elab{effdiff}
\frac{\partial \theta}{\partial t}=\kappa_\mathrm{eff} \nabla^2 \theta,
\eeq
for the large-scale concentration, with an \textit{effective diffusivity} $\kappa_\mathrm{eff}$ that accounts for the effect of the obstacles. This effect results from two competing mechanisms: obstacles reduce the area available to the scalar, which enhances dispersion, but  they also reduce the scalar flux, which inhibits dispersion. The second mechanism is dominant so that $\kappa_\mathrm{eff} \le1$ (see e.g.\ Ch.\ 1  of \cite{Jikov_etal1994}). \JV{Note that $\kappa_\mathrm{eff}$ is a scalar only if the obstacle arrangement is sufficiently symmetric, as considered in this paper, but is more generally a tensor.}

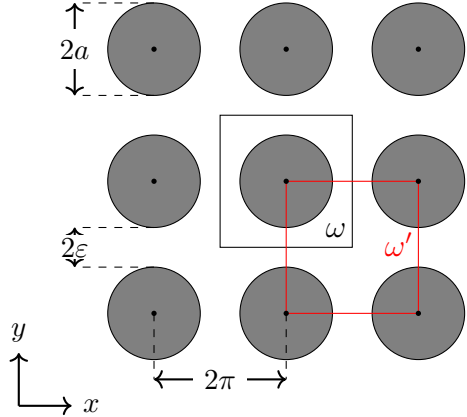
\begin{figure}
\centering
\begin{tikzpicture}
[nc/.style = {circle, draw=black, inner sep=0, minimum size=35, fill=gray},
every matrix/.style = {row sep={50,between origins}, column sep={50,between origins}, inner sep=0}]
\matrix (lattice) at (0,0) [] {
& \node(-1;-1)[nc]{}; & \node(-1;0)[nc]{}; &
\node(-1;1)[nc]{}; 
\\
&\node (0;-1)[nc]{}; & \node (0;0)[nc,label={[label distance =4pt]-47:$\omega$}]{}; & \node (0;1)[nc]{};
 \\
&\node [nc](1;-1) {};	& \node (1;0) [nc] {}; 	& \node(1;1) [nc,label={[label distance =4pt, red]95:$\omega '$}]{};
\\
};
\draw ($  (0;0)+(25pt,25pt) $) -- ($ (0;0)+(25pt,-25pt)$) -- ($ (0;0)+(-25pt,-25pt)$) -- ($ (0;0)+(-25pt,25pt) $) --cycle;
\draw [red]($(0;0)$) -- ($(0;1)$) -- ($(1;1)$) -- ($(1;0)$) --cycle;
\path
($  (-1;-1)+(0pt,17.5pt) $) edge [dashed] ($  (-1;-1)+(-30pt,17.5pt) $)
($  (-1;-1)+(0pt,-17.5pt) $) edge [dashed] ($  (-1;-1)+(-30pt,-17.5pt) $)

($  (0;-1)+(0pt,-17.5pt) $) edge [dashed] ($  (0;-1)+(-30pt,-17.5pt) $)
($  (1;-1)+(0pt,17.5pt) $) edge [dashed] ($  (1;-1)+(-30pt,17.5pt) $)

($  (1;-1)+(0pt,0pt) $) edge [dashed] ($  (1;-1)+(0pt,-25pt) $)
($  (1;0)+(0pt,0pt) $) edge [dashed] ($  (1;0)+(0pt,-25pt) $)
;
	  \draw [<->,thick]
	  ($  (-1;-1)+(-30pt,17.5pt) $)--node[yshift = 0pt, fill=white, inner sep=4pt, outer sep=0pt]{\small $2\radius$}($ (-1;-1)+(-30pt,-17.5pt)$);
	  \draw [<->,thick]
	  ($  (0;-1)+(-30pt,-17.5pt) $)--node[xshift = 0pt, fill=white, inner sep=1pt, outer sep=0pt]{\small $2\varepsilon$}($ (1;-1)+(-30pt,17.5pt)$);
	  \draw [<->,thick]
	  ($  (1;-1)+(0pt,-25pt) $)--node[yshift = 0pt, fill=white, inner sep=4pt, outer sep=0pt]{\small $2\pi$}($ (1;0)+(0pt,-25pt)$);
		\draw [black, thick, ->] (-85pt,-85pt) -- +(20pt,0) node [right] {\small $x$};
		\draw [black, thick, ->] (-85pt,-85pt) -- +(0,20pt) node [above] {\small $y$};
		\foreach \x in {-1,...,1}{
		  \foreach \y in {-1,...,1}{
			\draw  ($(\x;\y)$) node[fill=black,circle,inner sep=0pt, outer sep=0pt, text width=2pt] {};
			}}
\end{tikzpicture}
\caption{
Square lattice of circular obstacles  indicating the problem's geometric parameters and the
two alternative elementary cells $\omega$ and $\omega'$ used in the analysis. \JV{The  obstacles have radius   $a$ and are separated by gaps of width $2\varepsilon$}. }
\flab{problem_geometry}
\end{figure}

Homogenisation theory \cite[e.g.][]{Hornung1996,Torquato2002, MeiVernescu2010} provides a set of techniques for the computation of $\kappa_\mathrm{eff}$ that extends and systematises the approaches used by the early pioneers.  Explicit asymptotic results, valid when the obstacles occupy a small or large area fraction $\sigma$, are particularly valuable. For small area fraction -- the dilute limit -- Maxwell and Rayleigh's results \cite{Maxwell1873,Rayleigh1892} yield
\beq\elab{Maxwell}
\kappa_{\text{eff}}\sim 1-\sigma  \quad\textrm{as} \ \ \sigma\to 0,
\eeq
while for near-maximal area fractions -- the dense limit -- and for circular obstacles in the configuration of Fig.\ \fref{problem_geometry} Keller \cite{Keller1963} obtains
 \beq\elab{Keller}
 \kappa_{\text{eff}}\sim
\frac{2 (\pi/4-\sigma)^{1/2}}{\pi^{3/2}(1-\pi/4)}\quad\textrm{as} \ \ \sigma\to \pi/4.
  \eeq
Between these two limits, the value of $\kappa_\mathrm{eff}$ can be computed numerically  \cite{Perrins_etal1979}.

   The present paper is motivated by the recognition that, for the initial-value problem, the diffusion approximation \eqref{eqn:effdiff} predicted by homogenisation \JV{and the corresponding Gaussian distribution of the scalar concentration} have a limitation, specifically they apply only to the core of the scalar distribution, $|\bx-\bx_0| = O(\sqrt{t})$, and fail in the tails, $|\bx-\bx_0| \gg \sqrt{t}$. 
\JV{This mirrors a corresponding limitation of the central-limit theorem, which underpins homogenisation 
\cite[see e.g.][]{Jikov_etal1994,PavliotisStuart2007} and similarly does not apply to the tail probabilities.}
This limitation is particularly significant for applications in which low concentrations are critical, such as the migration of radioactive elements from underground nuclear water repositories which has been examined using homogenisation \cite{Allaire_etal2009,Amaziane_etal2010}. Our aim, therefore, is to develop a coarse graining of \eref{diff} that goes beyond homogenisation and captures the tails of the scalar distribution. This can be achieved by applying ideas of large-deviation theory \cite[e.g.][]{Touchette2009}, adapting the approach developed by \cite{HaynesVanneste2014a} for transport in periodic fluid flow to the diffusion with obstacles \eref{diff}.
 The approach, which we introduce in \S\ref{sec:ld}, provides an approximation to the concentration $\theta(\bx,t)$ that is valid for $|\bx-\bx_0| = O(t)$, thus improving on homogenisation; it requires the solution of a family of eigenvalue problems (see \eref{eval2} below)  which can be regarded as a generalisation of the cell problem \JV{that appears when homogenisation is used to compute  $\kappa_\mathrm{eff}$ \citep[e.g.][]{PavliotisStuart2007}.}
   The family of eigenvalue problems also determines the speed of propagation of certain reaction fronts \cite{Berestycki_etal2005} so our results are also useful in this context.

In \S\S\ref{sec:numerical}--\ref{sec:dense}, we focus on circular obstacles in the geometry of Fig.\ \fref{problem_geometry} and obtain explicit results demonstrating the value of the large-deviation approach.
We first solve the family of eigenvalue problems numerically for different obstacle area fractions $\sigma$ (\S\ref{sec:numerical}). The results show that diffusion with $\kappa_\mathrm{eff}$ 
\nw{provides} a satisfactory  approximation of the concentration tails only in the dilute limit $\sigma \to 0$; for general $\sigma$ and, most markedly in the dense limit $\sigma \to \pi/4$, the tails are much fatter than predicted by the diffusive approximation and display some anisotropy\JV{, unlike the Gaussian core of the distribution}. To explore this further, we examine the dense limit in detail in \S\ref{sec:dense}, where we develop an asymptotic theory which extends Keller's result \eref{Keller} to the large-deviation regime. This theory, based on a matched-asymptotics treatment of the large-deviation eigenvalue problems, recovers and subsumes a more straightforward extension, which replaces the continuous geometry by that of a network \cite{Berlyand2012} and captures part of the concentration tails. We assess the ranges of validity of the various approximations and test them against numerical solutions of the eigenvalue problems.

\section{Large-deviation approximation} \label{sec:ld}

  Our goal is to obtain an approximation for the concentration $\theta(\bx,t)$ for long times $t\gg 1$. The theory of large deviations \cite{FreidlinWentzell1984,Freidlin1985,Touchette2009} applied to periodic environments indicates that it takes
 the two-scale form   \cite{HaynesVanneste2014a,TzellaVanneste2016}  
\beq\elab{largedev}
\theta(\bx,t)=\phi (\bx,\bxi,   t) \, \e^{-t g(\bxi)},\quad \textrm{where} \ \  \bxi=\frac{\bx-\bx_0}{t}\in\mathbb{R}^2  
\eeq
and $\bx_0$ is the location of the initial scalar release, such that $\theta(\bx,0)=\delta(\bx-\bx_0)$\footnote{\JV{More general initial conditions can be treated by convolving this fundamental solution with the initial concentration.}}. 
Here $g$ is a rate (or Cram\'er) function  which provides a continuous approximation for the most rapid changes in $\theta$.
   It is non-negative, convex, and has a single minimum and zero located at $\bxi=\bm{0}$ that yields the maximum of $\theta$
 in the limit of $t\to \infty$.
 The (positive) correction term  $\phi$   (with $\ln\phi=o(t)$ as $t\to\infty$) has the same periodicity as the lattice,
\beq\elab{periodic}
\phi(\bx+\br_{m,n},\bxi,   t)=\phi(\bx,\bxi,   t),
\eeq
where $\br_{m,n}$ with $(m,\, n) \in \mathbb{Z}^2$ denotes the positions of the centroids of the obstacles.

Eq.\ \eref{largedev} introduces
the vector  $\bxi=(\xi,\eta)^\mathrm{T}$ defined on the whole of $\mathbb{R}^2$ which captures variations  on scales large compared with the size of the lattice cells. The vector
$\bx$, in contrast, is defined on the multiply-connected domain obtained by excising the obstacles from $\mathbb{R}^2$ and captures variations on the scale of single lattice cells.
The large separation between the two scales allows $\bx$ and $\bxi$ to be treated as independent.
Introducing \eref{largedev} into \eref{diff-eqn} then gives
\begin{subequations}\elab{diff-twovar}
\begin{align}\elab{diff-twovar-eqn}
\left(\nabla^2  -2 \nabla_{\bxi}g\cdot\nabla+|\nabla_{\bxi}g|^2-\bxi\cdot\nabla_{\bxi}g +g- \partial_t\right)\phi & \nonumber \\
+t^{-1}\left((\bxi+2 \nabla -2\nabla_{\bxi}g)\cdot\nabla_{\bxi}- \nabla^2_{\bxi}g \right)\phi
&+t^{-2}\nabla_{\bxi}^2\phi=0
\end{align}
while the boundary conditions \eref{noflux} become
\beq \elab{diff-twovar-bcs}
\bn\cdot(\nabla-\nabla_{\bxi} g+t^{-1}\nabla_{\bxi} )\phi=0 \quad \textrm{on} \ \ \mathcal{B}.
\eeq
\end{subequations}

Substituting the expansion
\beq\elab{twoscaleexp}
\phi(\bx,\bxi,t)=t^{-1}\left(\phi_0(\bx,\bxi)+t^{-1}\phi_1(\bx,\bxi)+t^{-2}\phi_2(\bx,\bxi)+\ldots\right),
\eeq
with the prefactor $t^{-1}$ motivated by mass conservation \cite{HaynesVanneste2014a}, into  \eref{diff-twovar} gives
 \begin{subequations}\elab{eval2}
 \beq
 \nabla^2\phi_0-2\bp\cdot \nabla \phi_0+|\bp|^2\phi_0=f(\bp)\phi_0 \elab{eval2a}
 \eeq
at leading order, where we have defined
\beq\elab{Legendre}
  \bp=(p,q)=\nabla_{\bxi} g(\bxi)\quad\text{and}\quad
  f(\bp) = \bxi\cdot \nabla_{\bxi} g(\bxi) - g(\bxi).
\eeq
The associated boundary conditions are deduced from \eref{diff-twovar-bcs} as
 \beq
 \bn\cdot (\nabla \phi_0-\bp  \phi_0)=0 \quad \textrm{on} \ \ \mathcal{B}.
 \elab{eval2b}
 \eeq
 \end{subequations}

Eqs.\ \eref{eval2}, together with the periodicity \eref{periodic} of $\phi_0$, define a family of eigenvalue problems parameterised by $\bp=(p,q)$, which determine a discrete spectrum of eigenvalues $f(\bp)$. The eigenfunctions can be thought \nw{of} as functions $\phi_0(\bx,\bp)$,  using the one-to-one correspondence between $\bxi$ and $\bp$.
The  eigenvalue problems can alternatively be rewritten in terms of
 \beq\elab{psi}
 \psi = \e^{- \bp \cdot \bx} \phi_0
 \eeq
 as the modified Helmholtz problems
  \beq
  \nabla^2\psi=f(\bp)\psi, \quad \bn\cdot\nabla \psi=0 \quad \textrm{on} \ \ \mathcal{B}, \quad \psi(\bx+\br_{n,m},\bp)=\e^{-\bp  \cdot \br_{n,m} }\psi(\bx,\bp),
  \elab{eval3}
  \eeq
involving Neumann conditions on the obstacle boundaries and a  `tilted' periodicity condition.

We focus on the principal eigenvalue $f(\bp)$  of \eref{eval2} or \eref{eval3}, that is, the eigenvalue with maximum real part,
with  associated eigenfunction $\phi_0(\bx,\bp)$ (unique up to multiplication).
The Krein--Rutman theorem implies that this eigenvalue is unique, simple and real. Moreover,
 $f\geq 0$ and convex.
The rate function $g$ is then deduced from $f$ by Legendre transform
since \eref{Legendre} together with convexity implies that $f(\bp)$ and $g(\bxi)$ are Legendre duals. Thus solving the family of eigenvalues problems \eref{eval2} or \eref{eval3} provide all the elements of the large-deviation approximation \eref{largedev} of the scalar concentration.

For small $|\bxi|$, the rate function can be  approximated as
 \beq\elab{g_quadratic}
 g(\bxi)\sim\frac{1}{2}\bxi^\mathbf{T} \nabla_{\bxi} \nabla_{\bxi} g(\bm{0}) \bxi
 =\frac{1}{4} \kappa^{-1}_{\text{eff}}|\bxi|^2,\quad |\bxi|\ll 1.
 \eeq
\JV{Here, the Hessian of $g$ at $\bm{0}$ is isotropic because we assume a four-fold symmetry for \eref{eval2}. For finite $|\bxi|$, $g(\bxi)$ is however not isotropic: dispersion is only approximately isotropic in the core of the scalar distribution and markedly anisotropic away from the core.}   
Introducing the  quadratic approximation \eref{g_quadratic} into \eref{largedev} recovers  the diffusive approximation for $\theta$ obtained via  
homogenisation of the diffusion equation \eref{diff}. 
\JV{The relationship between the large-deviation approach and homogenisation can be made completely explicit by noting that a perturbative solution of the eigenvalue problem 
\eref{eval2}   in the limit $|\bp|\to 0$ recovers, at leading order, the cell problem of homogenisation 
that determines $\kappa_\mathrm{eff}$. Pursuing the expansion to higher orders in $|\bp|$ yields corrections to $f(\bp)$ and $g(\bxi)$ that correspond to improvements to homogenisation involving diffusion-like operators of degrees higher than 2 \cite{MercerRoberts1990}. See \S 2.3 of \cite{HaynesVanneste2014a} for details of this expansion in the context of advection--diffusion.}

The eigenvalue problem \eref{eval2} or \eref{eval3} cannot be solved analytically, even for simple obstacle shapes. A useful lower bound can however be obtained: multiplying \eref{eval2a} by $\phi_0$, integrating by parts over an elementary cell $\omega$, and using the boundary   and periodicity conditions gives
\beq\elab{ineq}
f(\bp)\leq |\bp|^2 \quad\text{and}\quad  g(\bxi)\geq \frac{1}{4}|\bxi|^2
\eeq
with the second inequality obtained by taking a Legendre transform. Thus the presence of obstacles hinders dispersion
\cite[Th.\ 1.3]{Berestycki_etal2005} (note that $f$ and therefore $g$ may not vary monotonically with respect to the size of the obstacles \cite[Th.\ 1.4]{Berestycki_etal2005}).

We note that, owing to the close connection between large deviations and chemical-front propagation in the Fisher--Kolmogorov--Petrovsky--Piskunov (FKPP) model \cite[e.g.][]{Freidlin1985}, the principal eigenvalue of  \eref{eval2} or \eref{eval3} also determines the speed of these fronts  \cite{Berestycki_etal2005}, so that our results below also apply to this problem. \JV{We also emphasise that the eigenvalue problem \eref{eval2} or \eref{eval3}  
  readily generalises to more complex systems with periodic geometry including, for example, advection by a velocity field  \cite[e.g.][]{Mei1992,AuriaultAdler1995} and linear chemical reactions 
    \cite[e.g.][]{Mauri1991,AllaireHutridurga2012}. Large-deviation ideas are also potentially applicable to non-diffusive models of dispersion such as continous-time random walks which have been proposed for complex, non-periodic media  \cite[e.g.][]{Dentz_etal2018}.}  

\section{Circular obstacles in square lattices}\label{sec:numerical}

We now focus on the  simple geometry  of Fig.\  \fref{problem_geometry}, with circular obstacles of  radius $\radius$ arranged in square lattices with sides $2\pi$, so that
\beq\elab{r}
\br_{m,n}=2\pi
(m,n),
 \eeq
and the obstacle area fraction is $\sigma=\radius^2/(4\pi)$.
 We solve the eigenvalue problem \eref{eval2}  numerically using \nw{the} weak formulation described in Appendix \sref{weak}. We then use a standard finite-element discretisation: the eigenfunctions are approximated by continuous piecewise linear polynomials defined on a quasi-uniform triangular subdivision of the domain (obtained using Matlab’s PDE Toolbox). This results in a large, sparse generalised matrix eigenvalue problem that we solve using the Shift-and-Invert method \cite{Saad2011} to obtain an approximation for $f$ on a grid of values of $\bp$. Taking a numerical Legendre transform yields an approximation for $g$ as a function of $\bxi$.  We  now describe the results.

\begin{figure}
     \begin{center}
	 \begin{overpic}[width=.9\linewidth]{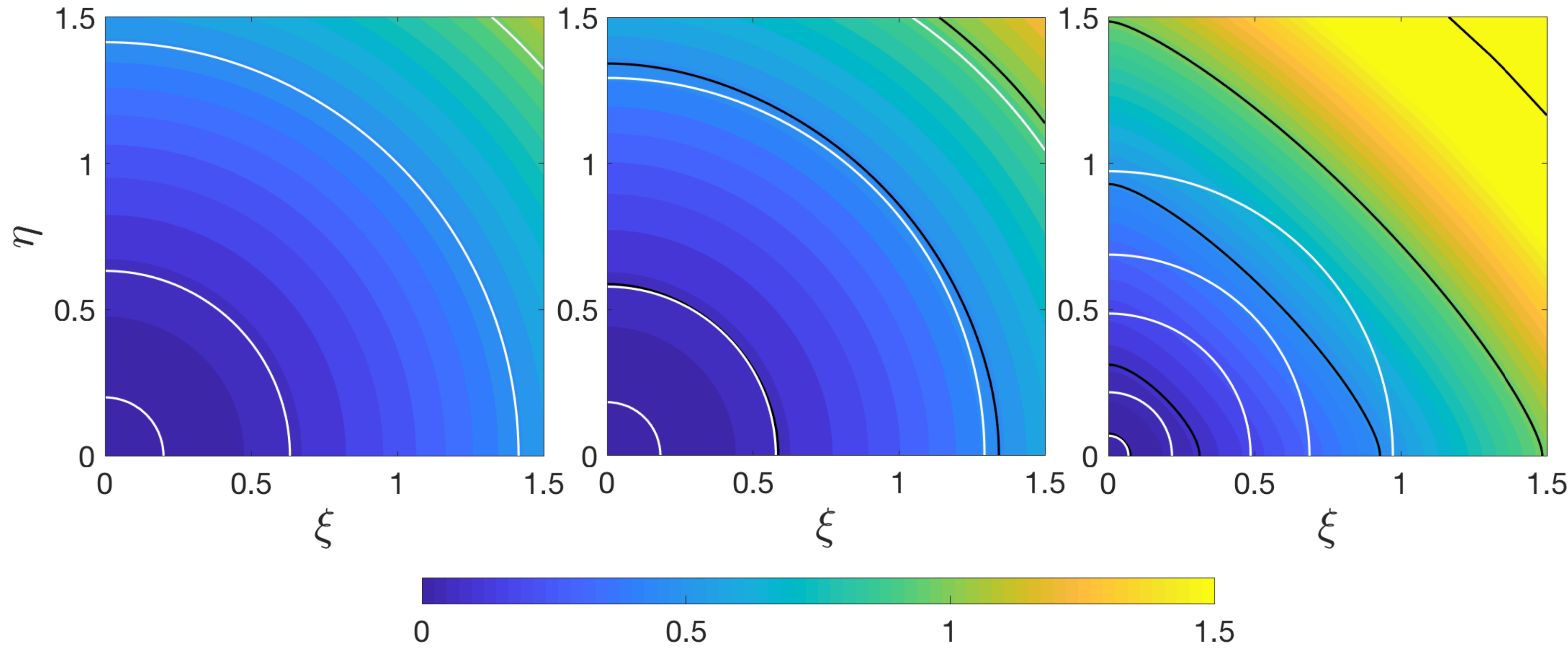}
     \end{overpic}
     \end{center}
  \caption{
  Rate function $g$ plotted  as a function of $\bxi$ for square lattices of circular obstacles with  radius
  (left) $\radius=0.01$, (middle) $\pi/2$
  and (right) $\pi-0.01$ (gap half width $\varepsilon=0.01$).
   Selected contours (with values $0.01$, $0.1$, $0.5$, $1$ and $2$) compare $g$ (black) with its quadratic \nw{(Gaussian)}  approximation (white)
	with the effective diffusivity $\kappa_\textrm{eff}$ given by  the Maxwell  formula \eref{Maxwell}  (left), a best-fit estimate (middle) and  the Keller formula \eref{Keller} (right).
            }
  \flab{g}
  \end{figure}

 Figure \fref{g} shows the rate function obtained for three obstacle radii $\radius=0.01, \, \pi/2$ and $\pi-0.01$ for $\bxi$ in the first quadrant (the other three quadrants are \nw{obtained from symmetry}).  The value $a=0.01$ is representative of the dilute limit $\radius\ll 1$, the value
  $a=\pi-0.01$ of the dense limit $ \pi-\radius \ll 1$,  
  with the area fraction close to the maximum  $\sigma=\pi/4$ allowed by the lattice arrangement.
 The figure  illustrates interesting features, such as anisotropy, which is absent for small obstacles (Fig.\ \fref{g}(a)) but very marked for the largest obstacles (Fig.\ \fref{g}(c)).
 As expected,  the quadratic  approximation  \eref{g_quadratic} of $g(\bxi)$ with effective diffusivity given by the Maxwell formula \eref{Maxwell} in the dilute limit (Fig. \fref{g}(a)) or  the Keller formula \eref{Keller} in the dense limit (Fig. \fref{g}(b)) is accurate near $\bxi=\bm{0}$.  In the intermediate case, the quadratic behaviour also holds near $\bxi=\bm{0}$ with an effective diffusivity that can be inferred from our results by contour fitting as an alternative to solving the cell problem of discretisation theory
 \cite{Perrins_etal1979}.
 Remarkably,
 in the dilute case, the quadratic approximation is  excellent
  beyond  the small-$\bxi$ neighbourhood and applies to the entire range of $\bxi$ shown.
In general and most strikingly
 in the dense case,
   $g$ is a more complicated function of $\bxi$
  for $|\bxi|=O(1)$.

The physical implications of the above results are that
homogenisation and  the corresponding  quadratic approximation  \eref{g_quadratic}  underestimate passive scalar transport.
The phenomenon
is most dramatic in the dense limit but negligible in the dilute limit.
This is illustrated in Figure \fref{logC} which focusses on two dense-limit cases:
  $\radius=\pi-0.01$ and $\pi-0.001$.  
  It shows  the (normalised) concentration $\theta(\bx,t)$ in logarithmic scale
along the diagonal $\bx=|\bx|(1,1)/\sqrt{2}$ obtained
 at five consecutive times
multiple of  $4\pi^2/\kappa_{\text{eff}}$. This choice ensures that
the time is sufficiently long for
the large-deviation approximation \eref{largedev} to apply.
 The figure   compares the   concentration
  obtained from the large-deviation approximation    with its  Gaussian, diffusive approximation \eref{g_quadratic} obtained   with effective diffusivity given by the Keller formula \eref{Keller}.
  Clearly
  the discrepancy between the large-deviation approximation and its  Gaussian, diffusive approximation is largest at early times, in the tails of $\theta(\bx,t)$ and for the largest of the two radii.
  As time increases, the Gaussian, diffusive approximation   describes  the bulk of the scalar concentration increasingly better.

\begin{figure}
     \begin{center}
	 \begin{overpic}[width=.8\linewidth]{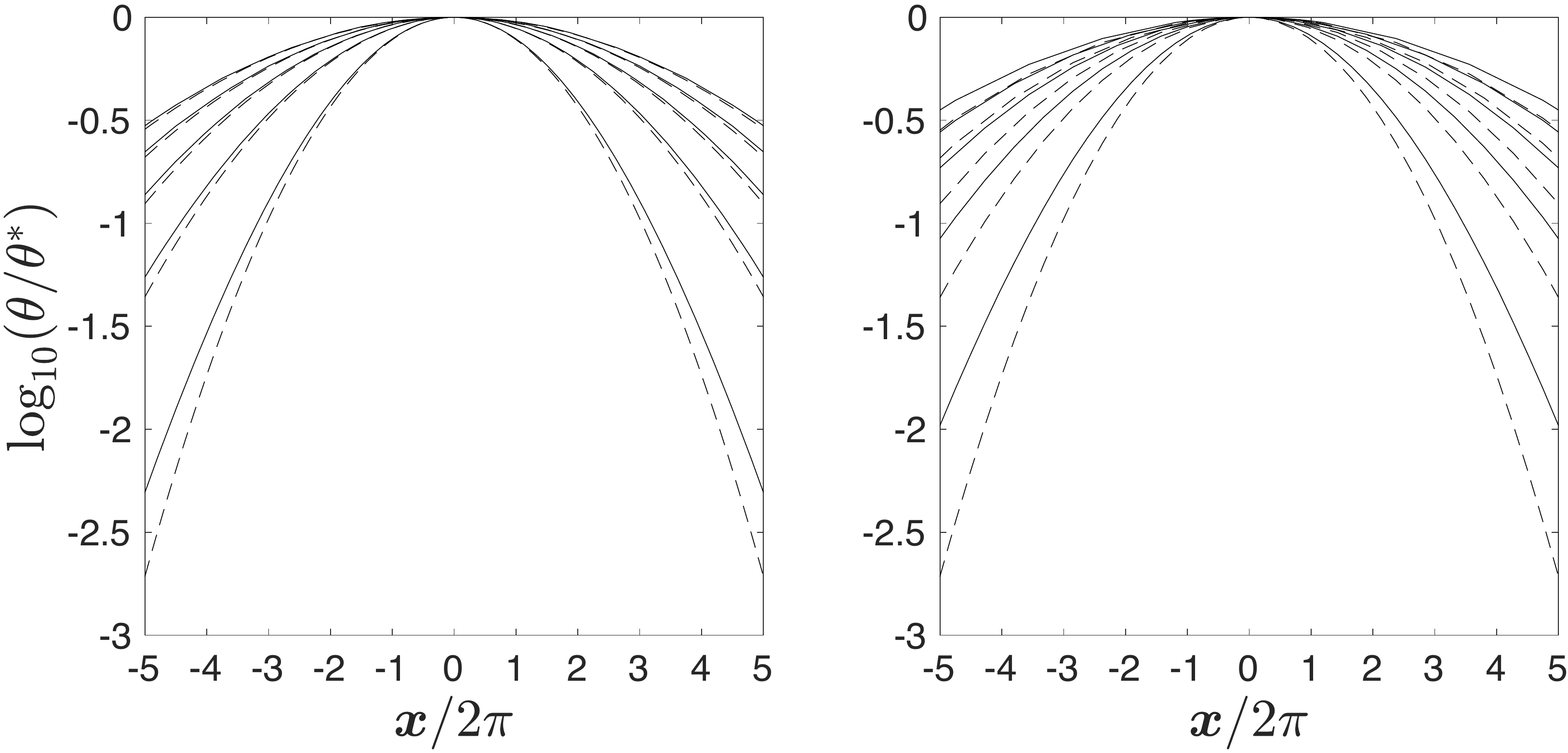}
 \linethickness{.5pt}
\put(36,36){\color{black}\vector(2.5,3){8.2}}
\linethickness{.5pt}
\put(87,36){\color{black}\vector(2.5,3){8.2}}
     \end{overpic}
     \end{center}
  \caption{
Normalised concentration $\theta(\bx,t)/\theta^{\ast}$  (in logarithmic scale) where $\theta^{\ast}=\text{max}_{\bx}\theta(\bx,t)$
 plotted against $\bx/(2\pi)$ for $\bx$ in the direction $(1,1)$ at dimensionless times $\kappa_{\text{eff}} t/(4\pi^2)=1$, $2$, $3$, $4$ and $5$, with the arrow pointing in the direction of increasing $t$,
for obstacles of radius $\radius=\pi-0.01$  (left)  and  $\pi-0.001$ (right)  (corresponding to gap half width $\varepsilon=0.01$ and $0.001$).
 Numerical results (solid lines)  obtained by solving \eref{eval2} are compared with  the Gaussian   approximation \nw{obtained from  \eref{effdiff}}  with effective diffusivity $\kappa_{\text{eff}}$ given by the Keller formula \eref{Keller} (dashed lines).
          }
  \flab{logC}
  \end{figure}

We can explain the validity of quadratic approximation  \eref{g_quadratic} with effective diffusivity $\kappa_\textrm{eff}$ given by the Maxwell formula \eref{Maxwell}
  throughout the range of $\bxi$ by solving the eigenvalue problem \eref{eval3} asymptotically in the dilute limit.
 This is carried out in  Appendix \sref{dilute} which shows that
  \beq \elab{fgdilute}
  f(\bp) \sim  \left(1 -  \sigma\right)|\bp|^2\quad
  \textrm{and} \quad
  g(\bxi)\sim  \tfrac{1}{4}\left(1 - \sigma\right)^{-1}|\bxi|^2
  \eeq
for $\sigma = {\radius^2 }/{4\pi} \ll 1$ and $|\bp|=O(1)$.
Thus, to leading order, the large-deviation approximation reproduces the results of classical homogenisation. In other words, the tails as well as the core of the distribution are Gaussian.

 In the remainder of the paper we focus on the dense limit where the (i) limitations of the diffusive approximation, (ii) non-Gaussianity and (iii) anisotropic behaviour are most prominent. We use the gap half
 width $\varepsilon=\pi-\radius \ll 1$  as small parameter for an asymptotic treatment.

\section{Dense limit} \slab{dense}

\subsection{Discrete-network approximation} \label{sec:network}

An intuitive way to understand the  dense limit is to consider
a discrete network model as a simplified analogue to the continuum model \eref{diff}. The building block of this model is Keller's asymptotic solution leading to \eref{Keller} \cite{Keller1963}. This relies on the observation that the scalar concentration $\theta(\bx,t)$ is nearly constant away from the small gaps separating neighbouring obstacles and changes rapidly along the gaps. The scalar flux is then localised in the gaps, unidirectional and approximately uniform in the direction transverse to the gaps. For a gap in the $x$-direction, for example, 
the total scalar flux  is given by
\beq \elab{KellerFlux}
F =2  h_\eps(x) \frac{\partial \theta}{\partial x},
\eeq
where $h_\eps(x)$    denotes the   gap
half width. For $\eps$ small, $h_\eps(x)$
can be approximated as a parabola centred in the middle of the gap: $h_\eps(x) \approx x^2/(2\pi)+\eps$.   Dividing across by $2h_\eps(x)$, integrating   and extending the integration range to $x \in (-\infty,\infty)$ gives the relationship
\beq\elab{totalflux}
F   = \alpha \Delta \theta, \quad \textrm{where} \ \ \alpha = \sqrt{2\eps/\pi^3},
\eeq
between the total scalar flux and the difference $\Delta \theta$ in concentration between the two sides of the gap.
This makes it possible to approximate \eref{diff}
by a discrete network in which regions away from the gaps are represented  as vertices and  the gaps between them as edges.
The  (near-uniform) concentration $\theta_{m,n}$  inside the region centred at $\pi(2m+1,2n-1)$
then evolves in response to the sum of the  fluxes in adjacent gaps, leading to
\beq\elab{networkmodel}
	\mathscr{A}
\frac{d\theta_{m,n}}{d t}=\alpha(\theta_{m+1,n}+
\theta_{m,n+1}+
\theta_{m-1,n}+
\theta_{m,n-1}-4 \theta_{m,n}),\ \
\eeq
where $\mathscr{A}=\pi^2(4-\pi)$
 is the approximate area of the region.
A  rigorous justification of model  \eref{networkmodel} can be obtained using the techniques in  \cite{Berlyand2012}.

   \begin{figure}
		\begin{center}
			 \begin{overpic}[width=.8\linewidth]{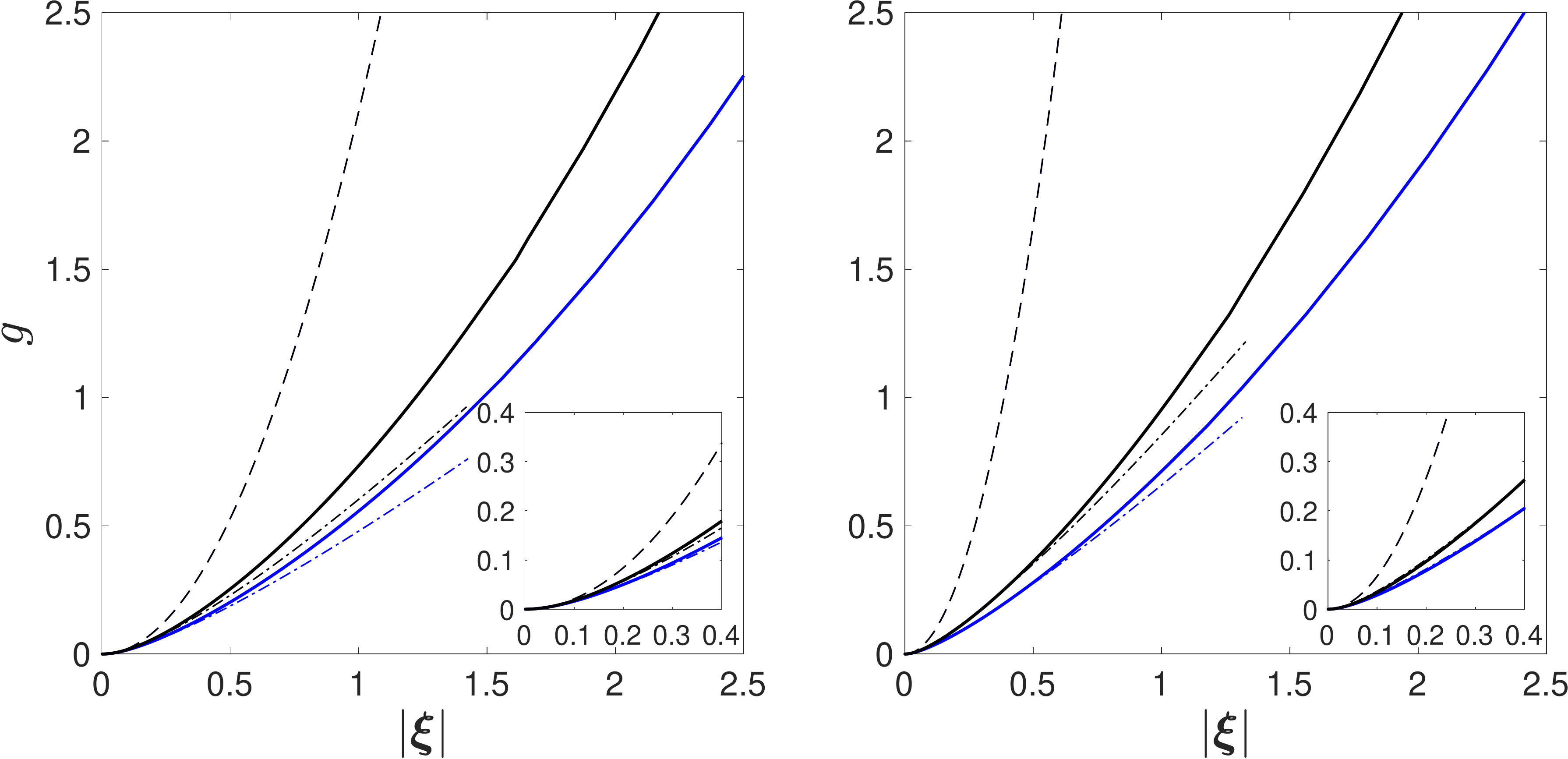}
     \end{overpic}
     \end{center}
    \caption{
   	Rate function $g$ plotted against $|\bxi|$
   	for obstacles of radius (left)  $\radius=\pi-0.01$  and (right)  $\pi-0.001$  (corresponding to gap half width $\varepsilon=0.01$ and $0.001$)
   	in the directions $(1,1)$ (black) and $(1,0)$ (blue).
    Numerical results (thick solid lines)  obtained by solving \eref{eval2} are compared
	with the quadratic \nw{(Gaussian)} approximation \eref{g_quadratic}
      (dashed lines)  \nw{with effective diffusivity $\kappa_{\text{eff}}$ given by the Keller formula \eref{Keller}} 
	   and the \nw{discrete}-network approximation \eref{g_network} (dashed-dotted).
	   The insets focus on small values of $|\bxi|$.
       }
    \flab{g_cross_dense}
    \end{figure}

It is easy to determine the long-time behaviour of \eref{networkmodel}.
 The diffusion approximation  is recovered by taking the continuum limit of   \eref{networkmodel}, approximating the right-hand side by $4 \pi^2 \alpha \nabla^2 \theta$ to obtain the effective diffusivity \beq
 \kappa_\mathrm{eff} = 4\pi^2 \alpha/\mathscr{A} = \alpha/(1-\pi/4)
 \eeq
 which is readily shown to match Keller's expression \eref{Keller} using that
 $\sigma = \pi/4 - \eps/2 + O(\eps^2)$.
 However,
 the diffusive approximation is limited.
Further information
 can be obtained
from  the rate function which appears
in the large-deviation approximation
   $\theta_{m,n}\sim t^{-1} \exp(-t g_\mathrm{d}(\br_{m,n}/t))$
of solutions of  the network model \eref{networkmodel}.
 Substituting  into \eref{networkmodel}  gives an explicit   relation for the Legendre transform  $f_\mathrm{d}(\bp)$ of $g_\mathrm{d}(\bxi)$, namely
\beq
f_\mathrm{d}(\bp)=
\frac{4\alpha}{\mathscr{A}}
\left(\sinh^2 (\pi p)+\sinh^2 (\pi q)\right).  \elab{eval_network}
 \eeq
 Taking the Legendre transform of \eref{eval_network} then yields
\begin{equation} \elab{g_network}
 g_\mathrm{d}(\bxi)=\frac{2\alpha}{\mathscr{A}}\left(\text{S}(\beta\xi)+\text{S} (\beta\eta)\right),
 \end{equation}
where $\text{S}(x) = 1+ x \sinh^{-1} x - \sqrt{1+x^2}$
and
$\beta=\mathscr{A}/(4\pi\alpha)$.
Figure \fref{g_cross_dense} shows that the rate function \eref{g_network}  is an improvement to the quadratic (Gaussian)  approximation.
Nevertheless  this improvement is  limited to small values of $|\bxi|$.  This is because a key assumption of the network model, namely that the concentration is nearly uniform outside the gaps, breaks down for $|\bx|$ large enough that $|\bxi|$ is not small. We now carry out an asymptotic analysis that yields an approximation to rate function that is valid for $|\bxi|=O(1)$ and recovers \eref{g_network}  as a limiting case.

\subsection{Asymptotics}

We apply matched asymptotics to obtain
an approximation
  to the solution of the eigenvalue problem \eref{eval3}  for $\eps \ll 1$  that is valid in the distinguished regime $|\bp|=O(1)$.
The analysis  is conveniently carried out
in  the translated elementary cell $\omega'$ shown in  Fig. \fref{problem_geometry}. This is centred on the star-like region which we will (inaccurately) refer to as `astroid' in the limit $\eps\to0$,
when the gaps close to form four cusps \nw{(see Fig. \fref{canonical2})}.

 We first consider the solution inside
 a representative
 gap, \nw{in the 
 the positive $x$-direction,  that for convenience we refer to as west of the   centre of $\omega'$.}
The gap boundaries   are given by
\beq\elab{circ_boundaries}
y=-\pi\pm h_{\varepsilon}(x), \quad\text{where}\quad h_{\varepsilon}(x)=\pi-\left((\pi-\varepsilon)^2-x^2\right)^{1/2},\quad 0\leq x\leq \pi-\varepsilon.
\eeq
This  form  suggests
an \emph{outer region} where $x$, $y+\pi=O(1)$
 in which case \eref{circ_boundaries}  may be approximated as the boundary of the astroid,
\beq\elab{boundary1}
y=-\pi\pm h_0(x)+O(\varepsilon)\quad\text{where}\quad h_0(x) = \pi-\sqrt{\pi^2-x^2},\quad 0\leq x\leq \pi,
\eeq
and an \emph{inner region} where $X = x/\sqrt{\varepsilon} = O(1)$ and $Y = (y + \pi)/\varepsilon = O(1)$ in terms of which
\eref{circ_boundaries} is approximately given by
\beq\elab{boundary2}
Y=\pm H_{\varepsilon}(X)=\pm H_0(X)+O(\varepsilon)\quad\text{where} \ \ H_0(X) = \frac{X^2}{2\pi}+1.
\eeq

\subsubsection{Inner region}\slab{inner_dense}

Inside the inner region, the modified Helmholtz problem \eref{eval3} becomes
\begin{subequations}\elab{eval2inner}
\beq
\varepsilon\partial_{XX}\Psi+\partial_{YY}\Psi =\varepsilon^2 f\Psi.
\elab{eval2ainner}
\eeq
for the eigenfunction $\Psi(X,Y)=\psi(x,y)$,
with the Neumann boundary condition approximated by
\beq
\pm\partial_Y\Psi-\varepsilon \left(\frac{X}{\pi}\partial_X\Psi
 \pm\frac{X^2}{2\pi^2}\partial_Y\Psi\right)+O(\varepsilon^{3/2})=0 \quad \textrm{at} \  \ Y =  \pm H_0(X)+O(\varepsilon).
\elab{eval2binner}
\eeq
\end{subequations}
We now introduce the expansions
\beq\elab{expansion_dense}
\quad\Psi = \Psi_0+\varepsilon^{1/2}\Psi_1+\varepsilon\Psi_2+O(\varepsilon^{3/2})
\quad\text{and}\quad f=f_0+\varepsilon^{1/2}f_1+\varepsilon f_2+O(\varepsilon^{3/2}).
  \eeq
of the eigenvalue and eigenfunction into \eref{eval2inner} to obtain, at $O(1)$ and $O(\varepsilon^{1/2})$,
\beq
		\partial_{YY} \Psi_i=0 \quad \textrm{with} \ \ \partial_Y \Psi_i=0\quad \textrm{at} \  \ Y = \pm  H_0(X)\quad (i=0,1)
\eeq
and thus  $\Psi_0=\Psi_0(X)$ and $\Psi_1=\Psi_1(X)$ i.e., they are transversely uniform. The key equation appears at $O(\eps)$. Using the $Y$-independence of $\Psi_0$ it simplifies to
\beq
		 \partial_{YY} \Psi_2=-\partial_{XX} \Psi_0  \quad \textrm{with} \ \  \mp\partial_Y \Psi_2 = \frac{1}{\pi} X \partial_X \Psi_0  \quad \textrm{at} \  \ Y =  \pm H_0(X). \elab{inlap}
\eeq
Integrating \eref{inlap} for $Y \in [-H_0(X), H_0(X)]$
 we find
that
$\partial_X (H_0(X) \partial_X \Psi_0) = 0$, hence
\begin{subequations}\elab{innersolutions}
\beq
\Psi_0 = A_\mathrm{1} \int_0^X \frac{\d X'}{h(X')} + B_\mathrm{1}=
\sqrt{2\pi}  A_\mathrm{1}  \tan^{-1} \left( \frac{X}{\sqrt{2\pi}} \right) +B_\mathrm{1},\quad X=x/\varepsilon^{1/2}. \elab{innersolutionsa}
\eeq
for constants $A_1$ and $B_1$ to be determined by matching with the outer solution.
 Similarly,
 using  symmetry, the solution inside the 
gaps to the south \nw{(in the negative  $y$-direction)}, east \nw{(in the negative  $x$-direction)},  and north \nw{(in the positive  $y$-direction)} of the centre 
\nw{of $\omega'$},  
are
\begin{align}
\Psi_0 &=
A_\mathrm{2}\sqrt{2\pi}\tan^{-1}\left(\frac{Y}{\sqrt{2\pi}}\right)+B_\mathrm{2}, \quad Y=(y+2\pi)/\varepsilon^{1/2},\elab{innersolutionsb}\\
\Psi_0 &=
A_\mathrm{3}\sqrt{2\pi}\tan^{-1}\left(\frac{X}{\sqrt{2\pi}}\right)+B_\mathrm{3},\quad X=(x-2\pi)/\varepsilon^{1/2},\elab{innersolutionsc}\\
\Psi_0 &=
A_\mathrm{4}\sqrt{2\pi}\tan^{-1}\left(\frac{Y}{\sqrt{2\pi}}\right)+B_\mathrm{4},\quad Y=y/\varepsilon^{1/2}, \elab{innersolutionsd}
\end{align}
\end{subequations}
 introducing additional constants $A_i$ and $B_i$ for $i=2,\, 3, \, 4$.
The constants are constrained by the `tilted' periodicity condition
 \eref{eval3}, giving
\beq\elab{constraint1}
(A_\mathrm{3},B_\mathrm{3})=\e^{-2\pi p} (A_\mathrm{1},B_\mathrm{1})
\quad\text{and}\quad
(A_\mathrm{4},B_\mathrm{4})=\e^{-2\pi q} (A_\mathrm{2},B_\mathrm{2}).
\eeq

\subsubsection{Outer region and matching}
In the outer region we  assume the expansion
\beq\elab{expansion_dense_outer}
 \psi = \psi_0+\varepsilon^{1/2}\psi_1+
 \varepsilon\psi_2
 +O(\varepsilon^{3/2}).
\eeq
To leading-order  $\psi_0$  satisfies
 \begin{subequations}\elab{dense_outer}
 \beq\elab{dense_outera}
 \nabla^2\psi_0=f_0\psi_0, \quad \bn\cdot\nabla\psi_0=0
 \quad\text{on}\quad y=-\pi\pm
 \left\{
 \begin{array}{cl}
	 h_0(x) & \textrm{for} \ \ 0 \le x < \pi \\
	 h_0(2\pi-x) &  \textrm{for} \ \ \pi \le x < 2 \pi
\end{array}
\right. .
 \eeq
Additional boundary conditions
  are obtained by matching the solution near the cusps of the astroid
  with  the inner solutions \eref{innersolutions}.
Near the cusp to the west of the centre, $\psi_0$ satisfies the approximation
$\partial_x(h_0(x)\partial_x\psi_0)=0$ to \eref{dense_outera}
(obtained following similar steps as in \S \sref{inner_dense}), hence
	\beq
	x^2 \partial_x\psi_0\sim C_1,\quad\text{as $\bx\to \bx_1=(0,-\pi)$},
\eeq
\end{subequations}
where $C_1$ is a constant to be determined.
The behaviour is similar near the other three cusps located
at  $\bx_2=(\pi,-2\pi)$, $\bx_3=(2\pi,-\pi)$ and $\bx_4=(\pi,0)$
 involving constants $C_i$ for $i=2,3,4$.

  \begin{figure}`
   \centerline{\includegraphics[width=0.7\linewidth]{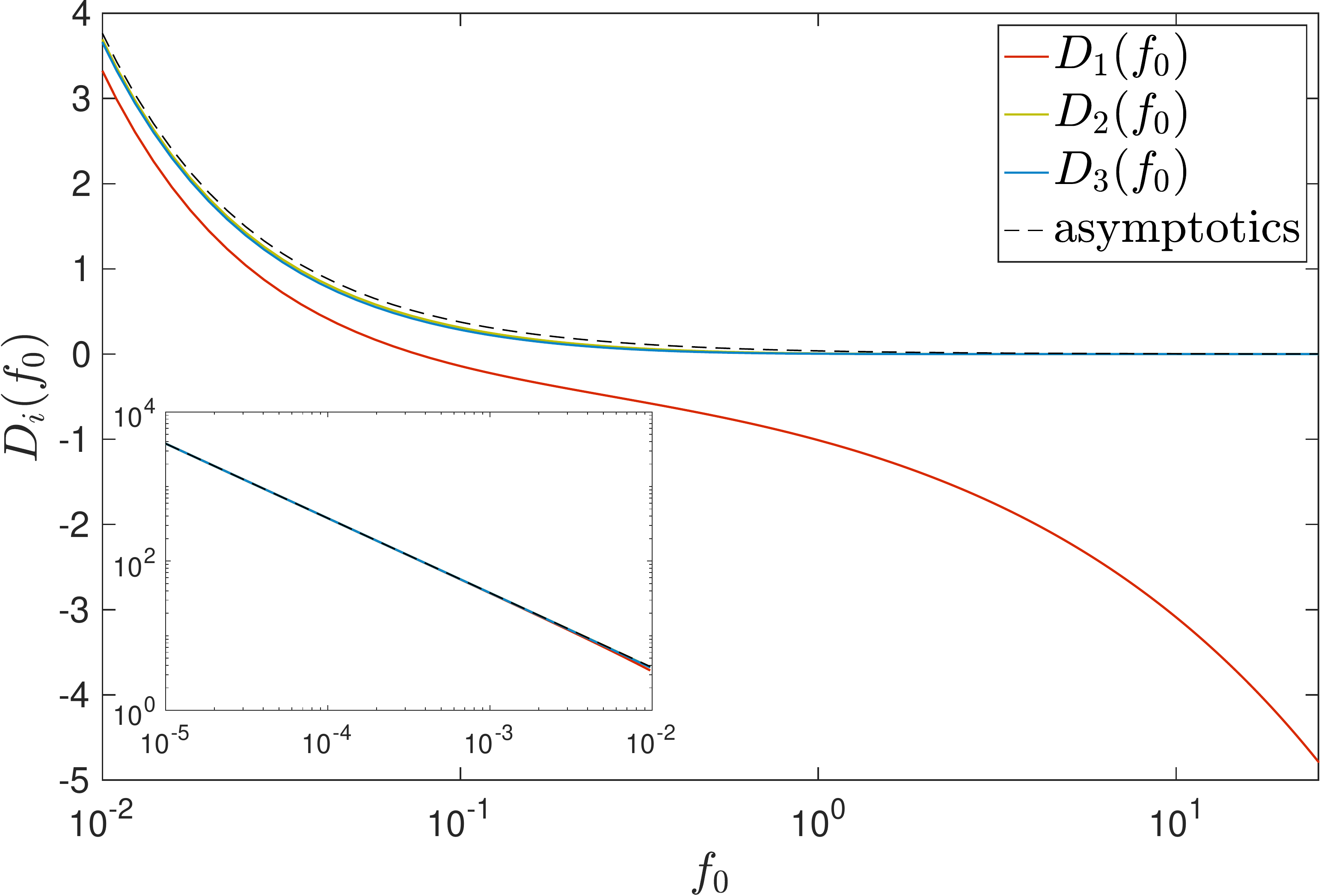}}
   \caption{
   Functions $D_i$ $(i=1,2,3)$ defined by \eref{outer_cusp}   against $f_0$.
   The numerical estimates (solid lines) are compared with  the asymptotic approximation \eref{D_smallf} for
   $f_0\ll 1$ (dashed line). }
   \flab{canonical1}
   \end{figure}

\paragraph{Canonical boundary-value problem.}
Exploiting the four-fold symmetry the solution to \eref{dense_outer}
can be written as the linear combination
	  \beq\elab{solution_outer}
	  \psi_0(x,y)=C_1\psi^{\ast}(x,y)+C_2\psi^{\ast}(y+2\pi,-x)+C_3\psi^{\ast}(2\pi-x,-y-2\pi)+
	  C_4\psi^{\ast}(-y,x)
	  \eeq
of the solution	 $\psi^{\ast}$ of
	 the canonical boundary-value    problem
	  \begin{subequations}\elab{canonical}
		   \beq\elab{canonicala}
	   \nabla^2\psi^{\ast}=f_0\psi^{\ast},
\eeq
with Neumann boundary conditions
on the astroid except on the western cusp where
\beq\elab{canonicalb}
x^2\partial_x \psi^{\ast}\to 1\quad\text{as $\bx\to \bx_1$}.
\eeq
\end{subequations}
The key point is that solving \eref{canonical} determines four functions
 $D_i(f_0)$, $(i=1,\ldots,4)$ defined by
\beq\elab{outer_cusp}
 \psi^{\ast}\sim -\frac{1}{x}-D_1(f_0)\ \ \textrm{as} \ \ \bx\to \bx_1 \quad \textrm{and} \quad
 \psi^{\ast}\sim -D_i(f_0)\quad\textrm{as} \ \ \bx\to \bx_i \ \ \textrm{for} \ \ i=2,\, 3,\, 4.
\eeq
Thus, these functions describe the leading-order behaviour of $\psi^{\ast}$ once the singular contribution $-1/x$ at $\bx_1$ is subtracted out.
By symmetry
 \beq\elab{constraint3}
 D_4(f_0)=D_2(f_0).
 \eeq

We obtain the  functions $D_i(f_0)$     by solving  \eref{canonical} numerically for a range of values of $f_0$ using  a standard finite-element discretisation.  
The difficulty associated with the singular shape of the astroid is avoided by trimming off the cusp regions with four straight segments placed a small distance $\delta$ away from  each cusp.
 Fig.\ \fref{canonical1} shows the results obtained for a range of values of $f_0$. These results have been checked to be insensitive to the value of $\delta$ as well as to the resolution of the finite-element discretisation ($\delta=0.01$ for the figure). Fig.\ \fref{canonical2} shows the form of $\psi^{\ast}$
 for different values of $f_0$.
 Clearly, larger values of $f_0$, corresponding to larger values of $\bp$, lead to higher contrasts in $\psi_0$, reflecting the fact that the asymptotic analysis goes beyond the hypothesis of near-uniform concentration assumed for the discrete-network approximation of  \S \ref{sec:network}.

  \begin{figure}
   \centerline{\includegraphics[width=0.8\linewidth]{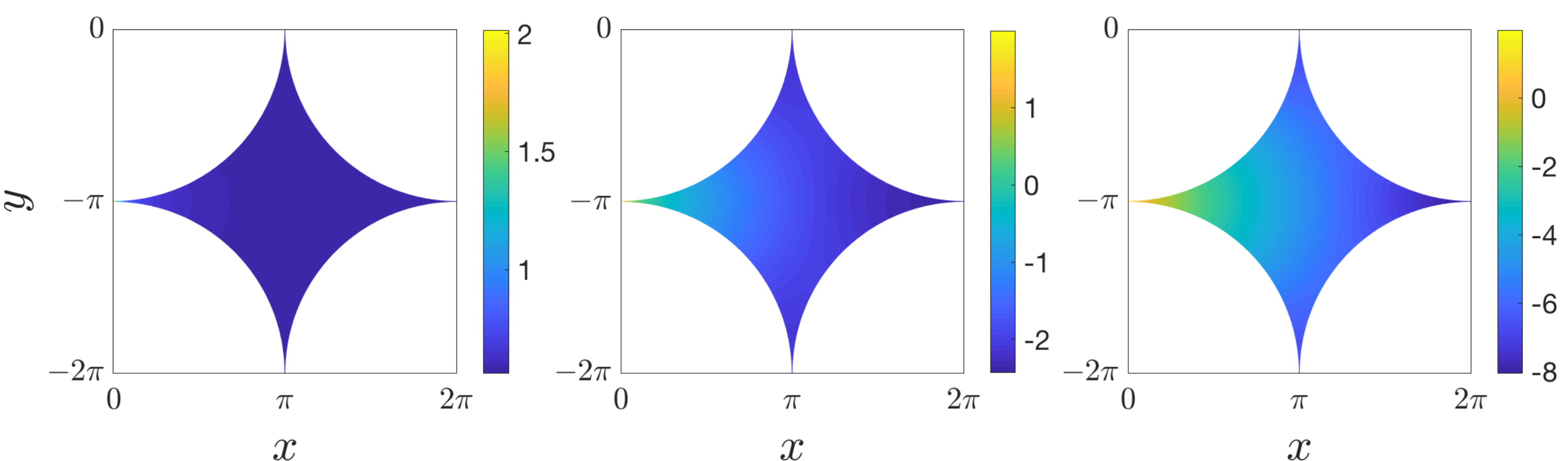}}
   \caption{
    Logarithm $\log_{10}|\psi^{\ast}|$ of
    $\psi^{\ast}(\bx)$
    defined as the solution
    to  the canonical boundary-value problem  \eref{canonical}
    obtained for (left) $f_0=0.01$, (middle) $1$ and (right) $10$.
	}
   \flab{canonical2}
   \end{figure}

 The asymptotic behaviour of $D_i(f_0)$ for small $f_0$ is useful for later reference. %
  In this limit,
  the solution $\psi^{\ast}$ may be expanded according to
  \beq\elab{canonical_outer}
  \psi^{\ast}=f_0^{-1}\psi^{\ast}_0+ \psi^{\ast}_1+O(f_0), \quad f_0\ll 1.
  \eeq
 Substituting \eref{canonical_outer} into \eref{canonicala}, we obtain that
   \begin{subequations}
   \beq\elab{psistarouter0}
   \nabla^2\psi^{\ast}_0=0\quad \text{and}\quad \nabla^2\psi^{\ast}_1=\psi^{\ast}_0.
   \eeq
 Additionally, from \eref{canonicalb} we have that
 $\psi^{\ast}_0$ and $\psi^{\ast}_1$ satisfy Neumann boundary conditions on the astroid except on the western cusp where $\psi^{\ast}_1$  satisfies
  \beq\elab{outer_cusp2}
      x^2\partial_x\psi^{\ast}_1\to 1\quad\text{as $\bx\to \bx_1$}.
  \eeq
    \end{subequations}
 Thus, $\psi^{\ast}_0=c$ where $c$ is a constant.
  The value of $c$ may be determined  by integrating the second equation in \eref{psistarouter0} and using \eref{outer_cusp2} to obtain
  \beq
 \mathscr{A} c=\int_{\omega'} \nabla^2 \psi^{\ast}_1 \d \bx \sim - \lim_{\delta \to 0} \int_{-\delta^2/(2\pi)}^{\delta^2/(2\pi)} \partial_x\psi^{\ast}_1|_{x=\delta}\,\d y=-\frac{1}{\pi},
  \eeq
 with the area $\mathscr{A}$ defined in \eref{networkmodel}.
 Therefore
  \beq\elab{D_smallf}
  D_i(f_0)\sim \frac{1}{\pi\mathscr{A}f_0}   \quad\text{for}\quad  f_0\ll 1 \quad (i=1,\cdots,4).
  \eeq
  Figure \fref{canonical1} confirms the validity of \eref{D_smallf}.

 \begin{figure}
      \begin{center}
 	 \begin{overpic}[width=.8\linewidth]{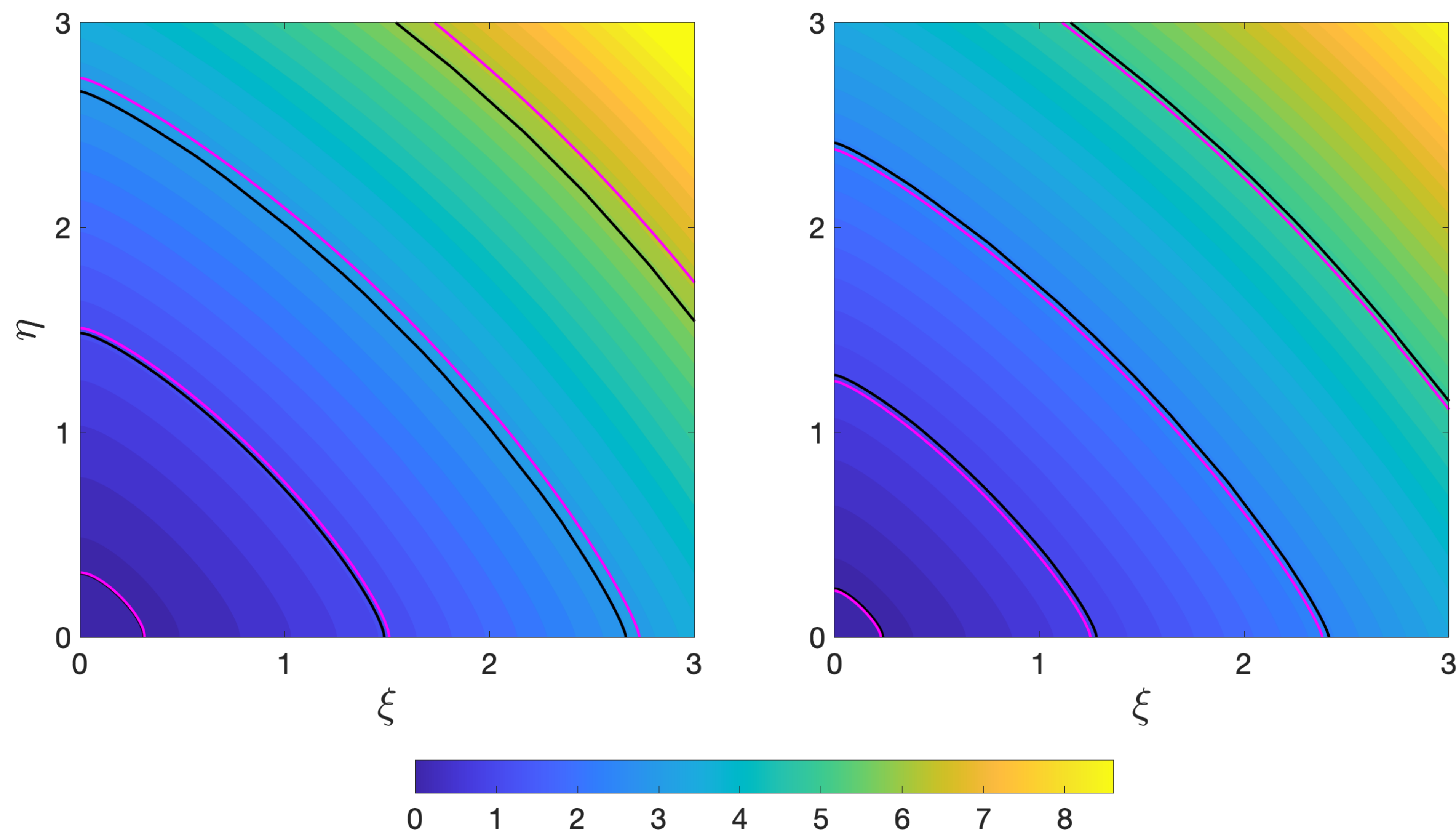}
      \end{overpic}
      \end{center}
   \caption{
  	Rate function $g$ plotted against $|\bxi|$
    	for square lattice of circular obstacles with radius $\radius=\pi-0.01$ (left) and  $\pi-0.001$ (right) (corresponding to gap half widths $\varepsilon=0.01$ and $0.001$) obtained by solving the eigenvalue problem \eref{eval2}.
    Selected contours (with values $0.1$, $1$, $2.5$ and $5$) compare the numerical  $g$  (black) with the asymptotic approximation deduced from \eref{transcedental} (pink).
           }
   \flab{g_dense}
   \end{figure}

\paragraph{Matching.} The leading-order approximation to the eigenvalue $f(\bp)$ is obtained by matching the solution in the inner and outer regions.   Comparing \eref{innersolutions} with \eref{solution_outer}  using
 \eref{outer_cusp} and \eref{constraint3} leads to
 \begin{subequations}\elab{matchrelations_full}
 \begin{align}
 C_1&=-2\pi\sqrt{\varepsilon}A_1,\quad
 &-C_1D_1(f_0)-(C_2+C_4)D_2(f_0)-C_3D_3(f_0)&=\sqrt{{\pi^3}/{2}}A_1+B_1,
 \elab{matchrelations_fulla}
 \\
 C_2&=-2\pi\sqrt{\varepsilon}A_2,\quad
 &-(C_1+C_3)D_2(f_0)-C_2D_1(f_0)-C_3D_4(f_0)&=\sqrt{{\pi^3}/{2}}A_2+B_2,\\
 C_3&=2\pi\sqrt{\varepsilon}A_3,\quad
 &-C_1D_3(f_0)-(C_2+C_4)D_2(f_0)-C_3D_1(f_0)&=-\sqrt{{\pi^3}/{2}}A_3+B_3,\\
 C_4&=2\pi\sqrt{\varepsilon}A_4,\quad
 &-(C_1+C_3)D_2(f_0)-C_2D_3(f_0)-C_4D_1(f_0)&=-\sqrt{{\pi^3}/{2}}A_4+B_4.
 \end{align}
 \end{subequations}
 We now use \eref{constraint1} to reduce \eref{matchrelations_full} to a homogeneous linear system for two of the constants, e.g., $A_1$ and $A_2$. Non-trivial solutions exist provided that the determinant of the associated matrix vanish. After some manipulations using \eref{constraint3} this gives
\begin{align}\elab{transcedental}
\left(D_3(f)\cosh(2\pi p)-D_1(f)-(2\pi\alpha)^{-1}\right) &
\left(D_3(f)\cosh(2\pi q)-D_1(f)-(2\pi \alpha)^{-1}\right) \nonumber \\
&-D_2^2(f)
\left(\cosh(2\pi p) -1\right) \left(\cosh(2\pi q)-1\right) = 0
\end{align}
with $\alpha=\sqrt{2 \eps/\pi^3}$ as defined in \eref{totalflux} and we omit the subscript of $f_0$.

Eq.\ \eref{transcedental} is the central result of this paper. It is a transcendental equation for the Legendre transform $f$ of the rate function $g$ as a function of the   gap half width $\eps$. Once the functions $D_i(f)$ have been tabulated, it reduces the determination $f$ and hence $g$ to an algebraic problem. The transcendental dependence of $f$ on $\eps$ reflects the uniform validity of our approximation across a range of values of $\bp$, including in particular a regime where $\sqrt{\eps} \e^{2\pi |\bp|}=O(1)$ as well as the discrete-network regime of \S\ref{sec:network}.

We solve \eref{transcedental} numerically for a range of $\bp$ to obtain $f(\bp)$ and $g(\bxi)$ by Legendre transform. In practice, it is convenient to express $\bp$ in polar form and, for fixed angle $\varphi$, solve for $|\bp|$ as a function of $f$ using a nonlinear solver such as Matlab's \texttt{fzero}.
The computation needs a good first guess which we obtain by noting that, when $\varphi=0$, i.e.\ for $\bp=|\bp|(1,0)$,  \eref{transcedental} reduces to $|\bp|=1/(2\pi)\cosh^{-1}((D_1(f)+\alpha^{-1})/D_3(f))$. We then  iterate over increasing values of $\varphi$ using the value of $|\bp|$ determined previously   as an initial guess for the next solution.

 \begin{figure}
	\begin{center}
	 \begin{overpic}[width=.8\linewidth]{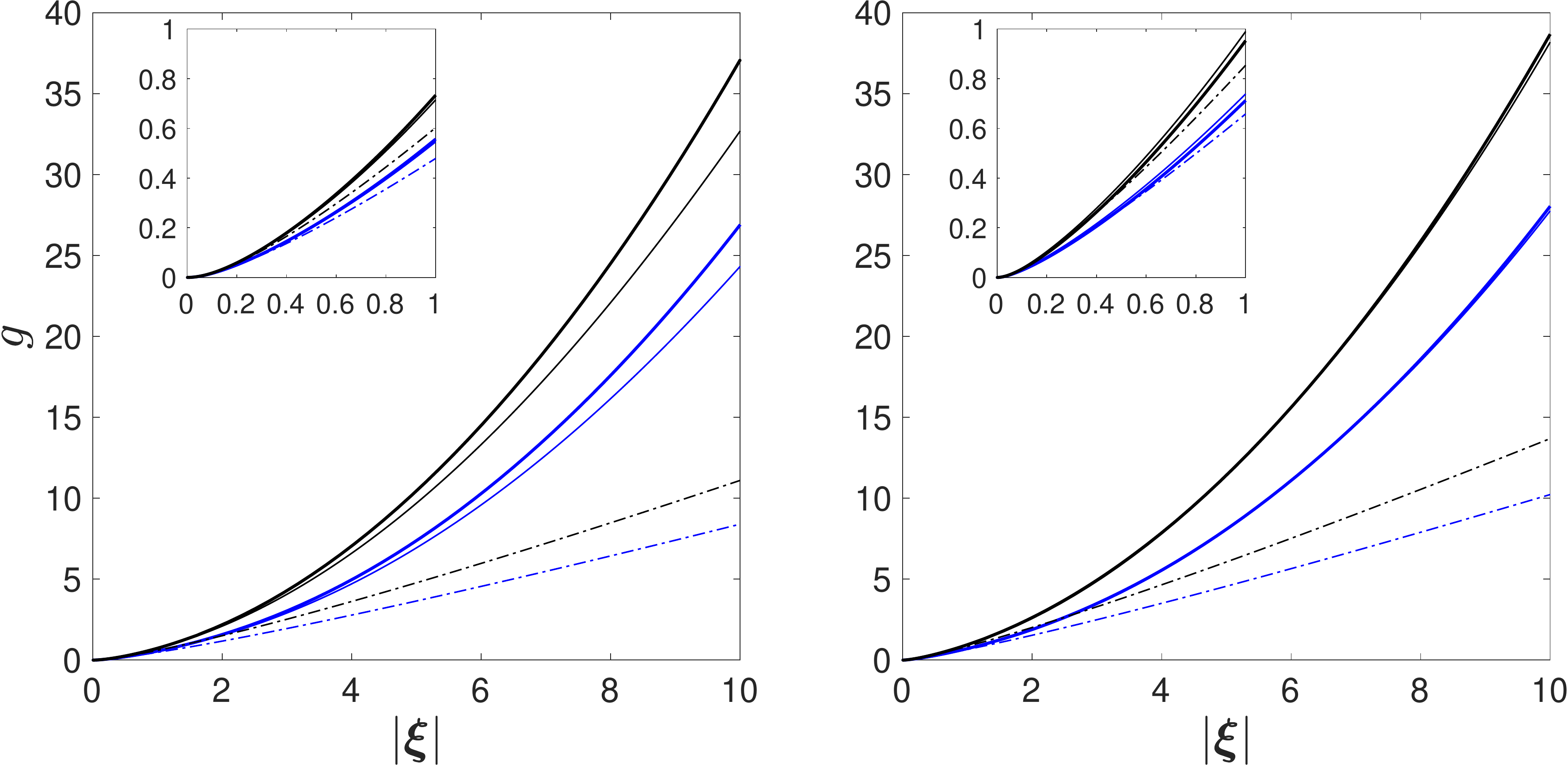}
    \end{overpic}
    \end{center}
   \caption{
   Cross-section of the rate function $g$ in Fig.\ \fref{g_dense} in the
	  directions $(1,1)$ (black lines)  and $(1,0)$ (lines).
   Numerical results  obtained by solving \eref{eval2} (thick solid lines)  are compared against the asymptotic approximation derived from \eref{transcedental} (thin solid lines) and from the discrete-network approximation \eref{eval_network} (dashed-dotted).
   The insets focus on small values of $|\bxi|$.
      }
   \flab{g_cross_dense_cross}
   \end{figure}

Fig. \fref{g_dense} compares the asymptotic prediction for $g(\bxi)$ deduced from \eref{transcedental} to that obtained by finite-element solution of the full eigenvalue problem \eref{eval2} for  $\varepsilon =0.01$ and $0.001$. The agreement is excellent throughout the range of $|\bxi|$, showing that our approximation captures the scalar concentration deep into the tails. This is more clearly demonstrated in Fig.\ \fref{g_cross_dense_cross} which displays cross-sections of $g(\bxi)$ for $\bxi$ along the $\xi$-axis and along the diagonal $\xi=\eta$ and for a very wide range of values of $|\bxi|$. While some discrepancies between asymptotic and numerical solutions are visible for $\eps=0.01$ the solutions match perfectly for $\eps=0.001$.

For very large $|\bxi|$, the concentrations are exceedingly small, of course. It is nonetheless interesting to note that $g(\bxi)$ is then controlled by an action-minimising trajectory as predicted by the  Friedlin--Wentzell (small-noise) large-deviation theory \cite{FreidlinWentzell1984}. This gives the asymptotics $g(\bxi) \sim d^2(\bxi)/4$ and hence $\theta \propto \e^{- d^2(\bx-\bx_0)/(4 t)}$ where $d(\bx)= \pi (x+y)/4+(1-\pi/4) |x-y|$ is the distance along the shortest path (made up of quarter circles and a line segment (horizontal if $x>y$, vertical if $x<y$)
joining   $\bx_0$ to $\bx$ while avoiding the obstacles.
Thus, at very large distances, one recovers a diffusive behaviour with the molecular value of the diffusivity but a non-Euclidean distance determined by the obstacle geometry (see \cite{TzellaVanneste2016} for a similar phenomenon in a different geometry).

We conclude by checking explicitly that our asymptotic analysis recovers the discrete-network approximation \eref{eval_network}. This approximation arises from the  transcendental equation  \eref{transcedental} in the limit of small $f$: introducing the small-$f$ asymptotic approximation \eref{D_smallf} of the $D_i$  into \eref{transcedental} gives
\begin{align}
\left(\cosh(2\pi p)-1 - \mathscr{A} f/(2\alpha)\right)& \left(\cosh(2\pi q)-1 - \mathscr{A} f/(2\alpha)\right) \nonumber \\
&- \left(\cosh(2\pi p)-1\right) \left(\cosh(2\pi q)-1\right)=0,
\end{align}
which simplies as $f = f_\mathrm{d}$ with the network expression \eref{eval_network} of $f_\mathrm{d}$. The corresponding rate function $g_\mathrm{d}$ is compared with the asymptotic and numerical estimates of $g$ in Fig.\  \fref{g_cross_dense_cross} which demonstrates the superiority of our asymptotic result over the network approximation.

\section{Conclusions}

This paper revisits the classical results of Maxwell, Rayleigh, Keller and many others since on the impact of an array of obstacles on the diffusion of scalars in an otherwise homogeneous medium.  Homogenisation theory predicts that, at a coarse-grained level, a scalar released instantaneously simply diffuses with a (computable) effective diffusivity.  A basic observation is that this conclusion is restricted to the bulk of the scalar distribution and that the more general tool of large-deviation theory is necessary to capture the tails of the distribution. Focusing on the case of a square lattice of circular obstacles for simplicity, we show that the non-diffusive behaviour is leading to tail concentrations that are much fatter than predicted by the effective diffusion  approximation. This effect is strongest in the dense limit, when the obstacles are nearly touching and large-scale dispersion is strongly inhibited. We examine this limit using a matched-asymptotics approach which reduces the computation of the large-deviation rate function -- requiring in general the solution of a family of elliptic eigenvalue problems -- to the algebraic equation \eref{transcedental}. The rate function that is obtained in this way captures the scalar concentration over a wide range of distances from the point of release and encompasses several physical regimes: the
diffusive regime with Keller's effective diffusivity \cite{Keller1963}, a closely related regime associated with a lattice random walk, all the way to the extreme-tail regime where the scalar concentration is controlled by single shortest-distance paths.

Our results exemplify a general phenomenon, relevant to a broad range of applications in porous media, composites and metamaterials, which takes its full significance when low concentrations are critical. This is the case in the presence of chemical reactions, as the example of the FKPP model makes plain. This adds the logistic term $\alpha \theta (1 - \theta)$, with $\alpha$ the reaction rate, to the right-hand side of the diffusion equation \eref{diff-eqn}. It leads to the propagation of fronts with speed $c(\bs{e})$ in the direction of the unit vector $\bs{e}$. This speed can be deduced from the rate function by solving $g(c(\bs{e}) \bs{e}) = \alpha$ or equivalently from its Legendre transform as $c(\bs{e}) = \inf_{p>0} \left(f(p \bs{e}\right) + \alpha)/p$ \cite{Freidlin1985,TzellaVanneste2015}. This provides an explicit example of a macroscopic manifestation of the tail behaviour of the scalar concentration.

We conclude by mentioning several directions in which our results could be extended.
 A first direction is to  adapt our   approach to consider different, more complex obstacle geometries, with extension from the square lattice to other Bravais lattices and from two to three dimensions, e.g.\ with spherical obstacles.
 A second direction would incorporate the effect of a \nw{steady}, incompressible flow. 
  \nw{The impact of the flow on dispersion is  determined by solving the appropriate   family of eigenvalue problems which include advection by a specified velocity field. This velocity field, driven for instance by a large-scale pressure gradient, can be obtained as part of the solution of a homogenisation problem posed for a fluid model such as the Stokes or Navier--Stokes models.}   
One expects that the  interaction between the inhomogeneity in the  flow and in the domain
will  lead to
interesting dispersion regimes,
dependent on the size of the obstacles and the intensity of the flow   (see e.g.\  \cite{Mei1992,AuriaultAdler1995} for the corresponding  \nw{homogenisation} regimes \nw{in the case of a   Stokes flow}). 
  A third direction concerns the nearly periodic case, introducing modulations in the  arrangement and size of the obstacles over long spatial scales  (see  \cite{BrunaChapman2015} for  corresponding homogenisation results).
    A fourth direction is to examine  random distributions of obstacles  such as those considered in   \cite{Jikov_etal1994,Torquato2002} \JV{or models of dispersion in complex media more sophisticated than simple diffusion \cite{Dentz_etal2018,MunicchiIcardi2020}.}
The fifth and final direction we suggest is the transfer of the tools of large-deviation theory from the (parabolic) diffusion equation to the (hyperbolic) wave equation, with applications to acoustics and photonics.
 Results are available
 about the effective wave speed (the analogue of the effective diffusivity)   in   media with obstacles, including in the dense limit \cite{Vanel_etal2017};
 it would be desirable to extend these to capture wave propagation over very large distances and to apply large deviations to go beyond simple dispersive corrections to homogenisation \cite{AbdullePouchon2018}.

\medskip

\noindent
\textbf{Data accessibility.}
The MATLAB code used to solve   \eref{eval2} numerically can be found in Bitbucket at
\url{https://bitbucket.org/loghind/eigfem-homogenisation/}.

\medskip

\noindent
\textbf{Acknowledgments.}
A. Tzella and D. Loghin acknowledge A. Patel for his contribution to a related MSci  project. 
Y. Farah   acknowledges support from  a PhD scholarship 
 from the UK Engineering and Physical Sciences Research Council.

\appendix

 \section{Dilute limit}\slab{dilute}
 We employ matched asymptotic expansions to approximate the solution to \eref{eval2} for $\radius\ll 1$ in the distinguished regime $|\bp|=O(1)$.
  This is straightforward to achieve by
 decomposing the elementary cell $\omega$ shown in Fig. \fref{problem_geometry}
   into an \emph{outer region} where $r=|\bx|=O(1)$,
   and an \emph{inner region} where $R=|\bX|=r/\radius=O(1)$,
 and by expanding
 \beq\elab{f_dilute}
 f\sim |\bp|^2+o(\radius),\quad \psi\sim \psi_0 +o(\radius)\quad\text{and}\quad\Psi\sim \Psi_0
 +o(\radius)\quad\text{as $\radius\to 0$},
 \eeq
 for the eigenvalue and  eigenfunction in the outer and inner regions respectively.
 Clearly, $\psi_0=\e^{-\bp\cdot\bx}$ is the leading-order solution to \eref{eval3}.
 Substituting the inner expansion inside \eref{eval3}, we find that
 $\Psi_0$ satisfies Laplace's equation
   \begin{subequations}\elab{interior_dilute}
   \beq
   \frac{1}{R}\frac{\partial}{\partial R}\left(R\frac{\partial\Psi_0}{\partial R}
   \right)+
   \frac{1}{R^2}\frac{\partial^2\Psi_0}{\partial{\theta^2}}=0
   \eeq
 in polar coordinates $(R,\theta)$,  with  Neumann  condition  on the boundary of the obstacle
   \beq
   \frac{\partial\Psi_0}{\partial R}=0\quad\text{at} \quad R=1.
   \eeq
 Let $\bp=|\bp|(\cos\varphi,\sin\varphi)$.
 The matching of $\Psi_0$ with $\psi_0$ is ensured
 provided that \beq \Psi_0=1-\radius |\bp| R \cos(\theta-\varphi)
 \quad\text{as $R\to \infty$}.\eeq
 \end{subequations}
 The solution to \eref{interior_dilute} is then
   \beq\elab{inner_dilute} \Psi_0=1-\radius |\bp|(R+R^{-1})\cos(\theta-\varphi).\eeq

 The leading-order inner and outer solutions may be used
 to obtain a  higher-order correction to  the eigenvalue
   $f(\bp)$.
 Multiplying \eref{eval3} by $\exp(\bp \cdot \bx)$ and integrating by parts over $\omega$ gives
 \beq\elab{integral_dilute}
 \int_{\partial \omega} \e^{\bp \cdot \bx} \bs{n} \cdot \nabla \psi \, \d l + \int_{\partial \omega} \e^{\bp \cdot \bx} \bs{n} \cdot \bp \psi \, \d l = (f-|\bp|^2) \int_{\omega}  \e^{\bp \cdot \bx}\, \psi \, \d \bx.
 \eeq
 We use the `tilted' periodicity condition  \eref{eval3} to deduce that $\exp(\bp \cdot \bx) \psi$ and $\exp(\bp \cdot \bx) \nabla \psi$ are periodic and therefore
 contributions along opposing edges of the square boundary cancel. After applying the boundary condition
 \eref{eval2b}, we are left with
 \beq\elab{integral_dilute2}
 (f-|\bp|^2) \int_{\omega}  \e^{\bp \cdot \bx} \psi \, \d \bx =   \int_{\mathscr{C}_\radius} \e^{\bp \cdot \bx} \bs{n} \cdot \bs{p} \psi \, \d l
 \eeq
 where $\mathscr{C}_\radius$ denotes the circle of radius $\radius$ centred at the origin.
 We use $\psi \sim \psi_0 = \e^{-\bp \cdot \bx}$
 to approximate the left-hand side of \eref{integral_dilute2}
 and
 \beq\elab{psidilute}
 \psi|_{r=\radius}=\Psi|_{R=1}\sim \Psi_0(1,\theta;\radius)
 = 1 - 2\radius |\bp| \cos(\theta-\varphi)+o(\radius)
 \quad\text{and}\quad
 \e^{\bp \cdot \bx} = 1 + \radius |\bp|\cos(\theta-\varphi) + o(\radius)
 \eeq
 to approximate the  right-hand side of \eref{integral_dilute2}.
 Carrying out the integrations we obtain $f(\bp)$ and, after Legendre transform, $g(\bxi)$ as the quadratic functions \eref{fgdilute}.

 We note that the quadratic approximations
\eref{fgdilute}
hold in the dilute limit for obstacles of arbitrary shapes, because only the far-field, dipolar form of the inner solution matters at leading order.

\section{Weak form of the eigenvalue problem \eref{eval2}}\slab{weak}

A weak form of the eigenvalue problem \eref{eval2} is readily obtained
by considering
 $\varphi$ to be  a general function
(square-integrable, including its derivatives)
satisfying the  same periodicity condition \eref{periodic} as $\phi$.
Multiplying \eref{eval2a} by $\varphi$ and integrating by parts over the elementary cell $\omega$ gives
\beq\elab{weak1}
(f(\p)-\Abs{\p}^2)\int_\omega\phi\varphi \dd\x
=
	 \int_{\partial\omega}\n\cdot\nabla\phi\varphi\dd s
	 -\int_\omega(\nabla\phi \cdot\nabla\varphi+2\p\cdot\nabla\phi\varphi)\dd\x.
\eeq
We use the   periodicity condition  \eref{periodic} to deduce that  contributions along opposing edges of the square boundary cancel.
Upon application of  the boundary condition  \eref{eval2b}, equation  \eref{weak1} becomes
\beq \elab{weak}
(f(\p)-\Abs{\p}^2)\int_\omega\phi\varphi\dd\x
	 =-\int_\omega(\nabla\phi\cdot\nabla\varphi+\p\cdot(\nabla\phi\varphi-\nabla \varphi\phi))\dd\x,
\eeq
where we have used
$\int_\omega \p\cdot\nabla\phi\varphi\dd\x=\int_{\mathscr{C}_\radius} \p\cdot\n\phi\varphi\dd s-\int_\omega\p\cdot\nabla\varphi\phi\dd\x$
to simplify.
A standard Galerkin finite element projection of \eref{weak} onto the
space of   degree-one continuous   piecewise-linear polynomials
defined
on  a quasi-uniform triangular subdivision of $\omega$
 yields a generalised eigenvalue problem that we solve numerically.

\providecommand{\noopsort}[1]{}\providecommand{\singleletter}[1]{#1}%

\end{document}